# Small Neutrino Masses: Another Anthropic principle aspect?

*C Sivaram*

Indian Institute of Astrophysics, Bangalore - 560 034, India

Telephone: +91-80-2553 0672; Fax: +91-80-2553 4043

e-mail: sivaram@iiap.res.in

*Kenath Arun[1]*

Christ Junior College, Bangalore - 560 029, India

Telephone: +91-80-4012 9292; Fax: +91-80- 4012 9222

e-mail: kenath.arun@cjc.christcollege.edu

*Kiren O V*

St. Josephs Indian Composite PU College, Bangalore - 560 001, India

Telephone: +91-9008789746; Fax: +91-80- 4012 9222

e-mail: kiren.ov@gmail.com

---

[1] Corresponding Author

**Abstract:** This year's Physics Nobel prize for the discovery of neutrino oscillations which resolved the problem of the missing solar neutrinos and the atmospheric muon neutrinos implies that at least one of the three neutrino species has a tiny mass. The neutrino oscillations measure the mass difference squared, and the individual neutrino masses have yet to be accurately ascertained. Particle theory has so far not given a predictive picture for neutrino masses. Here we propose that the anthropic principle may be relevant, as it is frequently invoked to understand other aspects of the universe, including the precise values of fine structure constant or nuclear coupling constant or even the proton-electron mass ratio.

**1. Introduction**

The discovery of neutrino oscillations which resolved the problem of the missing solar neutrinos and the atmospheric muon neutrinos, imply that at least one of the three neutrino species has a tiny mass, possibly of the order of one or a few electron volts. What the neutrino oscillations measure is the mass difference squared, i.e. $\Delta m^2 = m_1^2 - m_2^2$, between two species 1 and 2. More precisely what is obtained is the product of $\Delta m^2$ and the mixing angle, i.e. $\Delta m^2 \sin^2 2\theta$.

For neutrinos of given energy $E$, the oscillation length scales as, $E/\Delta m^2$. $\Delta m^2$ is typically of the order $10^{-2} - 10^{-4} \text{eV}^2$. Independent cosmological evidence, for instance, from the Wilkinson Microwave Anisotropy Probe (WMAP) suggest that the sum total of masses (Beringer *et al.* 2012; Hinshaw *et al.* 2013; Sivaram & Sinha 1974) is about an electron volt. This is also suggested by double $\beta_-$ decay experiments and earlier tritium decay end point analysis also implies a few electron volts (for a review see (Bahcall 1989)). The low neutrino masses also imply that neutrinos would constitute only a small fraction of the dark matter.

2. Primordial Neutrinos

Now neutrinos are expected to have been produced profusely in the very initial stages of the universe, i.e. in the hot big bang. Similar to the microwave background which is the fossil remnant of the hot radiation (high energy radiation) which characterized the best dense phase of the early universe epochs (cooling with expansion) we also expect a fossil remnant of neutrinos which also now form a background with an estimated density of about $150 \, cm^{-3}$, per species, so that summed over all six species (neutrinos and anti neutrinos), we expect a fossil neutrino background with a number density of one thousand per cubic centimetre (Kolb & Turner 1990; Peebles 1971).

So if each neutrino had a mass of about even twenty electron volts, this would imply that the universe would have a present density much greater than the closure density given by, $\rho_c = \dfrac{3H_0^2}{8\pi G}$ and would have collapsed several billion years ago. Most definitely a universe where a neutrino had a fifty electron volt rest mass (still ten thousand times lower than the electron rest mass, hitherto the lightest known elementary particle) would not have had much chance to develop biological life, let alone have advanced forms of life.

3. Neutrino Mass and Anthropic Principle

Since at present we do not have a definitive understanding of neutrino rest masses (the standard model says it should be zero), one wonders whether there can be some anthropic principle requirement for low masses (which would ensure enough cosmic time for cognitive life forms). The anthropic principle is frequently invoked to understand why coupling constants of fundamental interactions (like the fine structure constant or nuclear coupling constant or even the proton-electron mass ratio) have the measured values they have. Even small variations (like 2% change in nuclear coupling or doubling the electron mass) would make the universe untenable for biological life. For a detailed account see (Tipler & Barrow 1986). The anthropic principle has also been invoked (Weinberg 1987) to explain the value of the cosmological constant, now dominating the universe as dark energy (for alternative viewpoint see (Sivaram 1999)).

As our current theories in particle physics (including string theory) allow for a plethora of possible viable models (universes), the anthropic principle is invoked to understand the actual values we measure (in our universe). We could now add the requirement of light neutrino masses to this list. There are models to understand light neutrino masses but these invoke very heavy right handed neutrino (not detected). Then again the question arises as to why should they be so heavy?

**4. Conclusion**

In short particle theory has so far not given a predictive picture for neutrino masses (just as we do not understand electron rest mass either). The anthropic principle may again be relevant. Moreover, there being three neutrino species (in addition to three massive leptons) is related to the six flavours of quarks. At least three generations are required for a CP violation to generate excess of baryons in the early universe. This again seems an anthropic requirement, there should be a net excess of baryons of one part in $10^9$ to avoid complete annihilation to radiation in the early universe.

**Reference:**

Bahcall, J., 1989, Neutrino Astrophysics, Cambridge Univ. Press: Cambridge

Beringer, J. *et al*., 2012, Phys. Rev. D. 86, 010001

Hinshaw, G. *et al*., 2013, Astrophys. J. Suppl. Ser. 208, 19

Kolb, E. & Turner, M., 1990, The Early Universe, (Addison Wesley: Massachusetts

Peebles, P. J., 1971, Physical Cosmology, Princeton Univ. Press: Princeton

Sivaram, C. & Sinha, K. P, 1974, Current Science 43, 165

Sivaram, C., 1999, Mod. Phys. Lett. A. 14, 2363

Tipler, F. J. & Barrow, J. D., 1986, The Anthropic Cosmological Principle, Oxford Univ. Press: Oxford

Weinberg, S., 1987, Phys. Rev. Lett. 59, 2607